\documentstyle[twoside,fleqn,espcrc2,epsf]{article}

\title{
\vspace{-3.1cm}
\begin{flushright}
       {\normalsize KUNS-1408}    \\[-0.2cm]
       {\normalsize HE(TH)~96/09}  \\[-0.2cm]
       {\normalsize August 1996}   \\
\end{flushright}
       \vspace{0.7cm}
Gribov Copy and Complex Phase of Chiral Determinant
\thanks{Talk presented by Y.\,Kikukawa at {\sl Lattice\,'96},
St.\,Louis, USA, 4--8 June 1996.}
}

\author{
T. Aoyama$^{\rm a}$ and Y. Kikukawa
\address{
Department of Physics, Kyoto University, Kyoto 606-01, Japan}
}

\begin{document}

\begin{abstract}
We calculate the complex phase of chiral determinant by
the vacuum overlap formula with configurations of two-dimensional
U(1) gauge field fixed in Landau and Laplacian gauge.
The complex phase fluctuates over the Gribov copies,
which appear in the process of Landau gauge fixing
and contain vortex-like singularities.
In the Laplacian gauge, the fluctuation can be reduced and
the phase can be determined uniquely. If it is used as a
preconditioning for Landau gauge fixing,
the most smooth configuration is obtained among the copies generated.
\end{abstract}

\maketitle

\section{Introduction}
For a smooth background gauge field, the vacuum overlap
formula\cite{olnn,twist,schwinger,notwave,majorana,2dtorus_bc}
gives a definite and well-defined method to
calculate the complex phase of chiral fermion determinant
in lattice regularization.
The complex phase is gauge noninvariant by this method.
Its variation under a gauge transformation was shown to give
the consistent anomaly\cite{olnn,shamir_anomaly,al,rs}.
In anomaly-free U(1) chiral gauge theories on the two-dimensional
torus, the complex phase by the vacuum overlap formula reproduces
the exact result in the continuum limit with
uniform gauge fields\cite{twist}.

For generic gauge fields on the lattice, however, it was argued by
several authors\cite{waveguide,domainwall-ol-gs,notwave,2dtorus_bc} that
the nature of gauge degrees of freedom could spoil the chiral
structure of the formula.
For the (D+1) dimensional system of Wilson fermion with kink mass
and the wave guide of
gauge interaction\cite{kaplan}, it was shown\cite{waveguide}
that the gauge fluctuation induces another light fermion mode
of opposite chirality at the boundary of the wave guide.
Such a light mode could give the contribution to the complex phase of
chiral determinant.

On the other hand,
it was shown\cite{2dtorus_bc}
that the singular gauge transformation which wipe
out the {\it large} uniform vector potential into the singular
one with delta function, breaks the gauge invariance of
chiral determinant.
This happens in the anomaly-free U(1) chiral gauge theories on
the two-dimensional torus in the continuum limit, provided that
the anti-periodic boundary condition is chosen for all fermions.
The vacuum overlap reproduces the result as well.
The authors of ref.~\cite{2dtorus_bc} claimed that
it is these singular gauge transformations which spoils the
chiral phase when averaged on the gauge orbit.

In this article, we report our results of numerical study
about the gauge dependence of chiral determinant in
two-dimensional U(1) chiral gauge theory.
First we present our result on the
gauge integration of chiral determinant. We choose a
{\it small} uniform vector potential as background.
For this value,
the chiral determinant in the continuum limit is gauge invariant
even under the singular gauge transformation mentioned
above\cite{2dtorus_bc}.
Next we calculate the complex phase of chiral determinant with
the gauge field configurations fixed in Landau gauge.
We examine the effect of the singularity associated
to the Gribov copy\cite{shamir_lat95,vortex}.
Finally, we examine the performance of Laplacian gauge fixing,
especially as a preconditioning for Landau gauge\cite{laplacian}.

We consider the anomaly free U(1) chiral gauge theory
with four right-handed fermions with unit charge and
one left-handed fermion with charge two, which we call
``11112'' model following ref.~\cite{2dtorus_bc}.
The boundary condition is anti-periodic
for all fermions. The lattice size is $L=8$ and $16$.
The {\it small} uniform vector potential (toron) is chosen as
$A_1 L/2\pi= 0.20$ and $A_2 L/2\pi= 0.15$.
Generic U(1) gauge fields are generated at the value of
the gauge coupling $ e L /\pi = 0.05$ by the Metropolis algorithm.

\section{Gauge Integration}
With the toron in consideration,
the complex phase of the determinant for the ``11112'' model
takes the following value on the lattice $L=16$,
\begin{equation}
  \label{chiral-phase-toron}
  \phi_0= 0.13304 .
\end{equation}
Gauge functions, $g(n)_k$ ($k=1,...,N$) are randomly generated and
the variations of the complex phases by the gauge transformations,
$\phi_i-\phi_0$ are calculated. Then the complex phase $\phi$ of the
gauge-integrated determinant is calculated as
\begin{equation}
  \label{phase-averaged}
  \exp ( \Delta S + i \phi )
%= \frac{1}{N} \sum_{k=1}^{N} \exp( i \phi_k) .
= \left\langle \exp( i \phi_k) \right\rangle.
\end{equation}
In Figure 1, we show a result for $N=15,000$. We plot the
averaged value for each bin of 750 gauge transformations and also
the accumulated average over 750 times the number of bins $I$.
\begin{figure}[htb]
\vspace{-16pt}
\epsfysize=5.5cm
\centerline{\epsffile{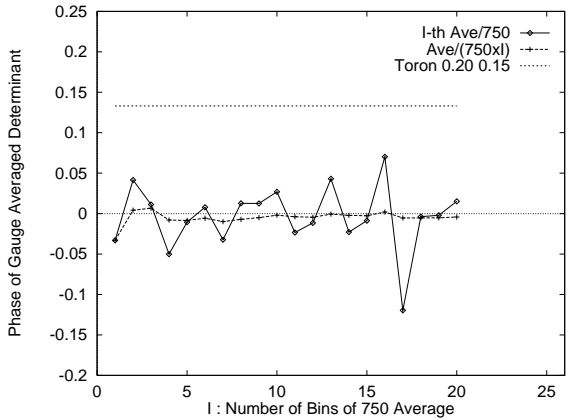}
}
\vspace{-24pt}
\caption{Complex phase of chiral determinant
averaged over $N=15,000$ gauge functions.}
\label{fig:gauge-integration}
\end{figure}

\vspace{-16pt}
\noindent
We repeated such measurement three times and obtained
\begin{equation}
  \label{phase-averaged-result}
\phi = -0.0035 \pm 0.006
\end{equation}
The standard error is used in this estimate.
The result shows
that the large gauge fluctuation affects
the chiral phase even for the {\it small} toron.

\section{Gauge Fixing and Chiral Determinant}

We first show, in Figure 2, a typical gauge potential
of the Gribov copy of the toron in consideration,
which is generated by the usual local algorithm for Landau gauge fixing.
The gauge function which generates such a copy from the original toron
is also shown. We can clearly observe the vortex singularity\cite{vortex}.

\begin{figure}[htb]
\vspace{-28pt}
\epsfysize=5.5cm
\centerline{\epsffile{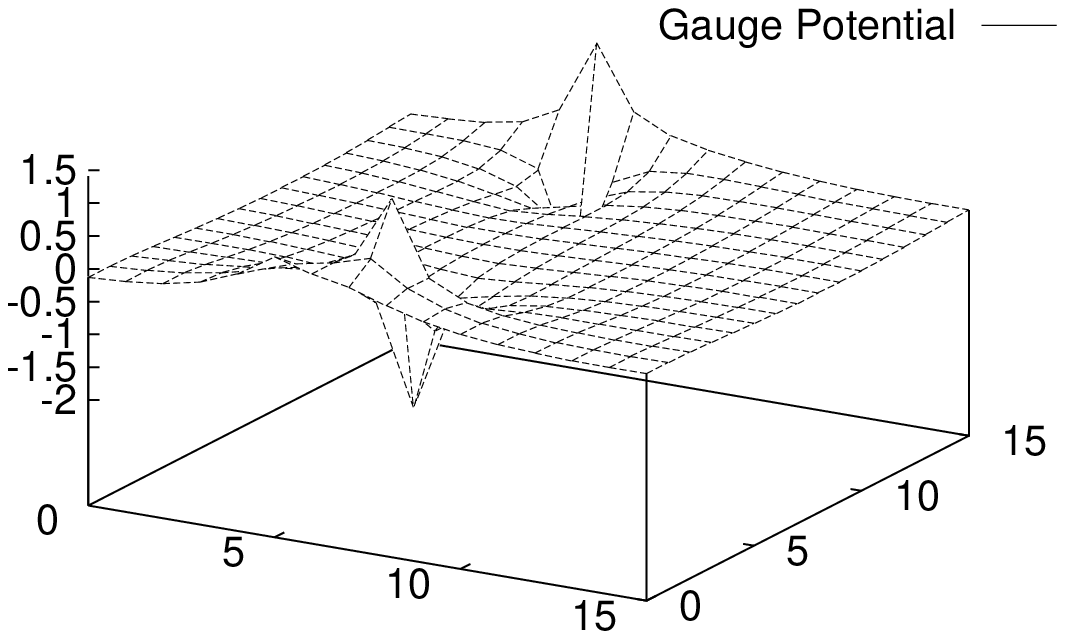}
}
\vspace{-16pt}
\epsfysize=5.5cm
\centerline{\epsffile{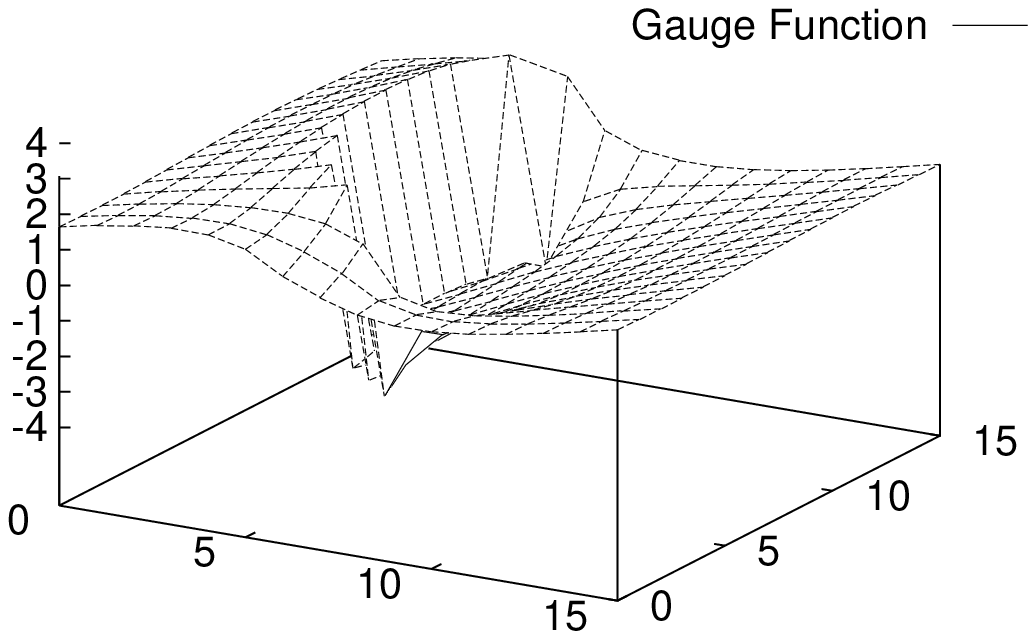}}
\vspace{-40pt}
\caption{Vector potential and gauge function of a Gribov copy
of the toron, $A_1 L /2\pi=0.2$, $A_1 L /2\pi=0.15$. $L=16$.}
\label{fig:gribov-copy-potential-gauge}
\end{figure}

\vspace{-16pt}
In Figure 3, we show the complex phases
calculated in Landau gauge
for a generic U(1) gauge field at $e L /\pi =0.05$.
We can observe the fluctuation of the complex phases.
It is clear that the singularity in the gauge function associated
to the Gribov copy affect the complex phase
very much\cite{shamir_lat95}.

The complex phases calculated in Laplacian gauge are also
shown in Figure 3. This gauge reduces the fluctuation of gauge degree
of freedom completely and gives a unique complex phase on the gauge
orbit.
The result of its use as the preconditioning is also impressive.
It gives the most smooth configuration among the Gribov copies
generated, as far as we examined. The obtained value of the complex
phase is also reasonable.

\begin{figure}[htb]
\vspace{9pt}
\epsfysize=5.5cm
\centerline{\epsffile{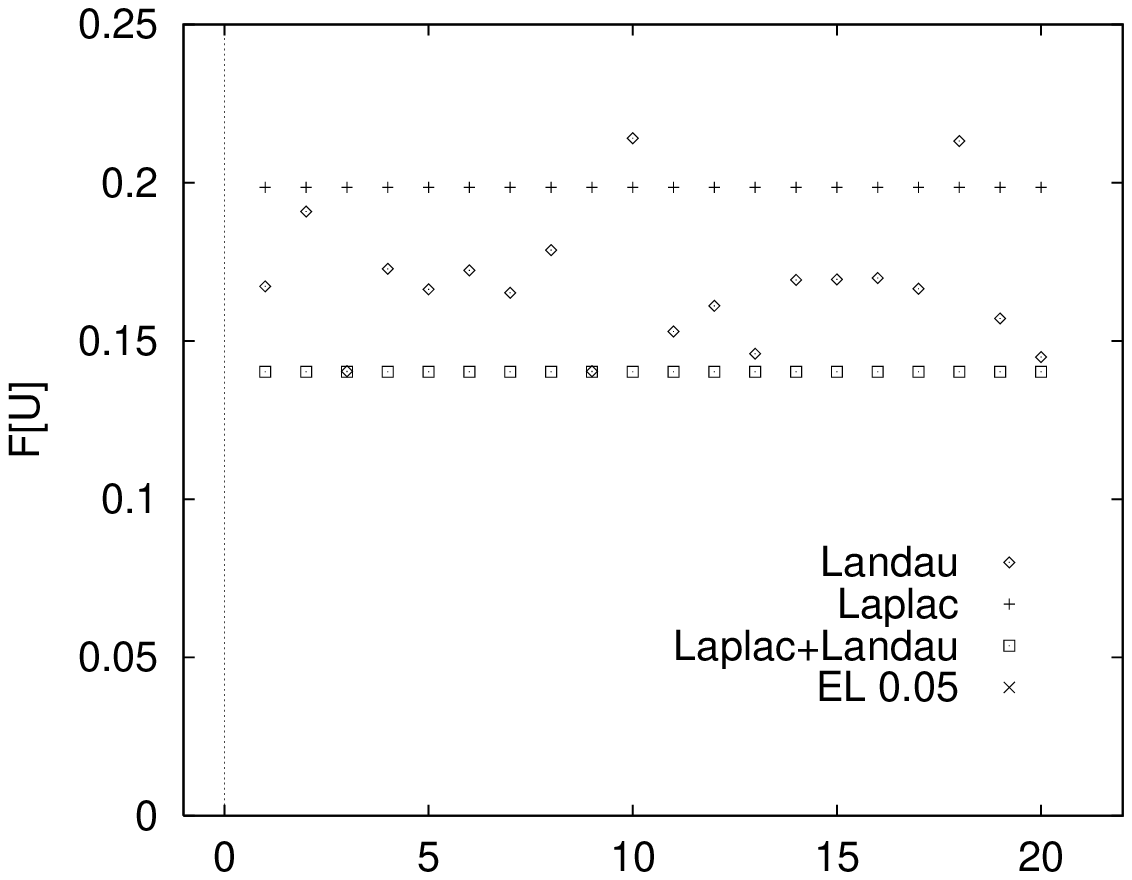}}
\vspace{9pt}
\epsfysize=5.5cm
\centerline{\epsffile{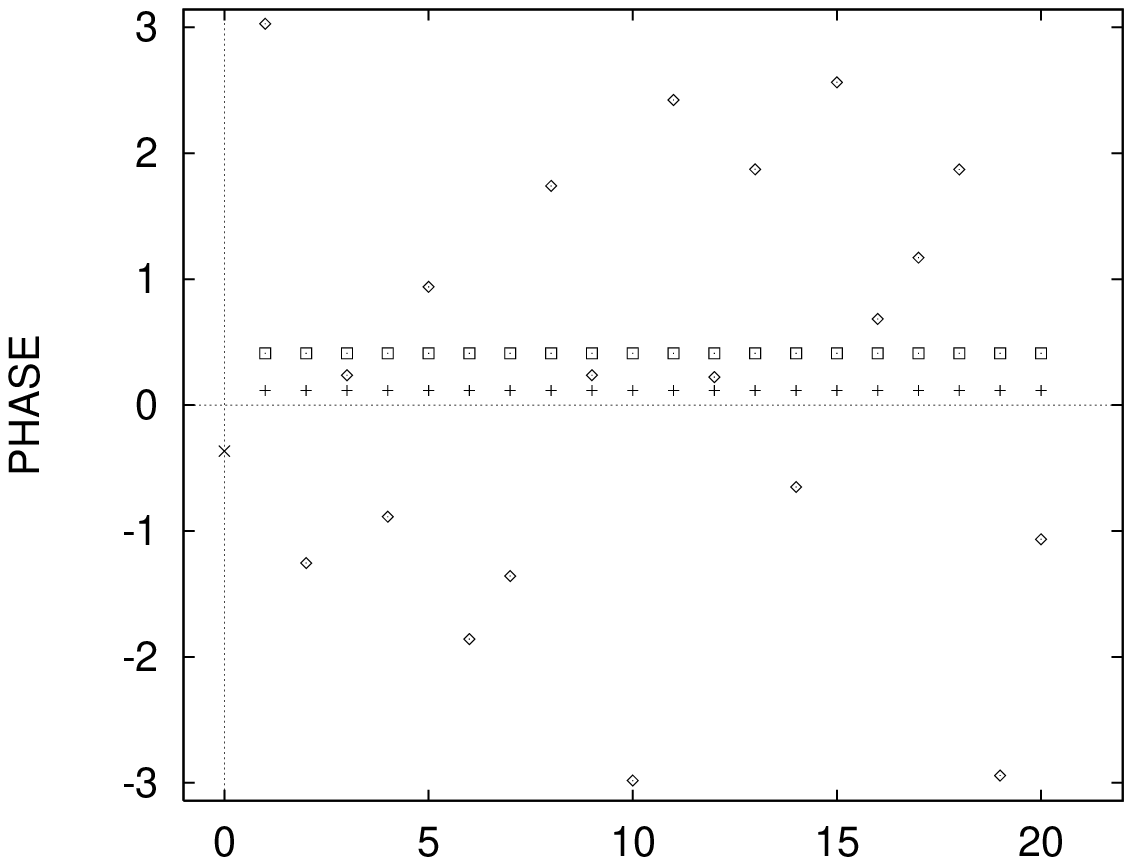}}
\vspace{-24pt}
\caption{Minimized values of $F[U]$ and
complex phases of chiral determinant
for Gribov copies generated from a generic electric field
at $eL/\pi=0.05$. 20 successive random gauge transformations
are applied. $L=8$. }
\label{fig:gribov-electric-fu-phase}
\end{figure}

\vspace{-40pt}
\section{Conclusion}
The complex phase defined by the vacuum overlap formula
is very sensitive to the singular gauge functions associated to
the Gribov copies.
It is hard to determine a unique phase on a
gauge orbit by gauge-fixing into Landau gauge with usual local
algorithm. The Laplacian gauge fixing reduces the gauge fluctuation
completely and gives a unique phase. If the gauge coupling constant
is small enough, the fixed configuration is rather smooth.
Preconditioning for Landau gauge fixing by the Laplacian method is
practically useful because, as far as we examined, it gives
the most smooth configurations among the Gribov copies.

This work is supported by National Laboratory for High Energy Physics,
as KEK Supercomputer Project (Project No. 10).

\end{document}